\documentclass[aps,prl,twocolumn,nofootinbib,preprintnumbers,amssymb,amsfonts,amsmath,superscriptaddress,showpacs]{revtex4-1}
\usepackage{graphicx}
\usepackage{bm}
\usepackage{hyperref}
\usepackage{color}
\usepackage{subfigure}

\def\apj{ApJ}

\def\apss{Ap\&SS}               
\def\planss{P\&SS}              

\def\jgr{Journal of Geophysical Research}

\def\simleq{\; \raise0.3ex\hbox{$<$\kern-0.75em \raise-1.1ex\hbox{$\sim$}}\; }
\def\simgeq{\; \raise0.3ex\hbox{$>$\kern-0.75em \raise-1.1ex\hbox{$\sim$}}\; }

\newcommand{\GeV}{{\rm GeV}}
\newcommand{\GV}{{\rm GV}}
\newcommand{\TeV}{{\rm TeV}}

\newcommand{\erg}{{\rm erg}}

\newcommand{\kpc}{{\rm kpc}}
\newcommand{\pc}{{\rm pc}}

\newcommand{\cm}{{\rm cm}}

\newcommand{\s}{{\rm s}}

\begin{document}

\title{Three-Dimensional Model of Cosmic-Ray Lepton Propagation Reproduces Data from the Alpha Magnetic Spectrometer on the International Space Station}

\author{Daniele Gaggero}
\email{dgaggero@sissa.it}
\affiliation{SISSA, via Bonomea 265, I-34136, Trieste, Italy}
\affiliation{INFN, sezione di Trieste, via Valerio 2, I-34127, Trieste, Italy}

\author{Luca Maccione}
\email{luca.maccione@lmu.de}
\affiliation{Ludwig-Maximilians-Universit\"{a}t, Theresienstra{\ss}e 37, D-80333 M\"{u}nchen, Germany}
\affiliation{Max-Planck-Institut f\"{u}r Physik (Werner Heisenberg Institut), F\"{o}hringer Ring 6, D-80805 M\"{u}nchen, Germany}

\author{Giuseppe Di Bernardo}
\email{giuseppe.dibernardo@physics.gu.se}
\affiliation{Department of Physics, University of Gothenburg, SE 412 96 Gothenburg, Sweden} 
\affiliation{Department of Astronomy and Theoretical Astrophysics Center, University of California Berkeley, Berkeley, California 94720, USA}

\author{Carmelo Evoli}
\email{carmelo.evoli@desy.de}
\affiliation{{II.} Institut f\"ur Theoretische Physik, Universit\"{a}t Hamburg, Luruper Chaussee 149, D-22761 Hamburg, Germany}

\author{Dario Grasso}
\email{dario.grasso@pi.infn.it}
\affiliation{INFN, sezione di Pisa, Largo B. Pontecorvo 3, I-56127 Pisa, Italy}

\preprint{LMU-ASC 25/13, MPP-2013-114}
\begin{abstract}
We study the compatibility of Alpha Magnetic Spectrometer (AMS-02) data on the cosmic-ray (CR) positron fraction with data 
on the CR electron and positron spectra provided by PAMELA and Fermi LAT. We do that in terms of a 
novel propagation model in which sources are distributed in spiral arm patterns in agreement with 
astrophysical observations. While former interpretations assumed an unrealistically steep injection 
spectrum for astrophysical background electrons, the enhanced energy losses experienced by CR 
leptons due to the larger average source distance from Earth allow us to reproduce the data with harder injection spectra as expected in a 
shock acceleration scenario. Moreover, we show that in this approach, and accounting for AMS-02 results, 
the contribution of nearby accelerators to the fluxes at very high energy can be significantly reduced, 
thus avoiding any tension with anisotropy upper limits. 
\end{abstract}

\maketitle 

{\em Introduction.} The detection of a rising positron fraction (PF) above 10 GeV, as measured by PAMELA \cite{Adriani:2008zr}, Fermi LAT \cite{Adriani:2011xv}, and Alpha Magnetic Spectrometer (AMS-02) \cite{Aguilar:2013qda}, is one of the most striking recent results in astroparticle physics. 
In the standard scenario, electrons are mainly accelerated in cosmic-ray (CR) sources, while positrons are only produced by collisions of CR protons and helium in the interstellar medium. The fraction of those secondary $e^+$ to the total $e^- + e^+$ (CRE) flux reaching the solar system should decrease with energy because of the decreasing escape time of the primary nuclei from the Galaxy as inferred by the observed secondary/primary nuclei ratios. 

Fermi LAT observations of a rising $e^+$ absolute spectrum \cite{FermiLAT:2011ab} prove the presence of an excess in the $e^{+}$ channel and, together with the CRE spectra \cite{Abdo:2009zk,Ackermann:2010ij} and the $e^{-}$ spectrum \cite{Adriani:2011xv} measured by Fermi LAT and PAMELA respectively, exclude that the rise of the PF is due to a steep $e^{-}$ spectrum.

The presence of an extra primary CR positron component at very high energy either of astrophysical or of exotic origin, seems therefore unavoidable.
Following this approach, several analyses succeeded in consistently reproducing the PF measured by PAMELA, the Fermi LAT CRE spectrum and other independent CR data sets (see \cite{Serpico:2011wg} for a comprehensive review). Most of them were based on the standard assumption, common to both semianalytical and numerical models, that CR sources are distributed smoothly in the Galaxy.

An important problem arising in this approach is that the required injection spectrum of the primary $e^{-}$ component is very steep: If a power-law spectrum is assumed according to standard acceleration theory, the spectral index lies in the range $2.6\div2.7$ depending on the details of the propagation model. This is significantly steeper than that inferred from radio observations of SNRs, $\langle\gamma\rangle=2.0\pm0.3$ \cite{Delahaye:2010ji}. Moreover, these values for the slope are quite different from the values $2.2-2.4$ required to reproduce the CR nuclei spectra. Therefore, it is very difficult to reconcile this scenario with shock acceleration theory which generally predicts the same spectral index, close to $2-2.3$, for electrons and nuclei \cite{Caprioli:2011ze}. 

However, the Galaxy has a spiral arm structure where also astrophysical CR sources are more likely to be found, while the solar system lies in an underdense region between two arms. Therefore, high energy CREs experience more energy losses than estimated within the standard scenario, due to the increased average distance they have to propagate through. This softens their observed spectrum, thus offering a realistic alternative to steep injection (see e.g.~\cite{Shaviv:2009bu}, although the model discussed there is in strong tension with the Fermi LAT $e^{+}$ spectrum). 

In this Letter we use a newly developed 3D propagation code ({\sc DRAGON.v3}~\cite{Dragonweb}) to account for the spiral arm distribution of CR astrophysical sources and prove the effectiveness of this mechanism. We will show, for the first time, that we can reproduce the observed spectra of CR species with a primary $e^{-}$ injection index close to the one used for nuclei and in rough agreement with radio observations of SNRs. We will assume that the extra-component sources have the same spatial distribution as the primary ones. This may be compatible with enhanced secondary production in aged SNRs, as suggested in \cite{Blasi:2009hv,Mertsch:2009ph}, or $e^{\pm}$ acceleration by pulsars \cite{Hooper:2008kg}. 

Furthermore, in our model a prominent contribution from local sources is not required. We avoid then any tension with two relevant features of the CRE data: (a) the smoothness of the observed CRE spectra, confirmed also with high accuracy by AMS-02, in contrast to the bumpiness expected if several local sources contribute to the high-energy fluxes; (b) the stringent upper limits placed by Fermi LAT and AMS-02 (in this case for the PF only) on the CRE dipole anisotropy, which are already challenging, albeit not excluding yet, the local source scenario \cite{Ackermann:2010ip,DiBernardo:2010is}. Remarkably, AMS-02 results suggest the extra component to be softer than inferred from PAMELA observations \cite{Cholis:2013psa,Yuan:2013eba,Yin:2013_pulsars}. This makes the relative contribution of local sources, if any, less relevant than previous analyses suggested. We note that, although several SNRs and pulsars are observed in the nearby region ($d <$~few hundred pc), none, or only a few, of these sources, may significantly contribute to the observed $e^\pm$ flux, either because they are not powerful enough or because propagation may take place along streams which have a small probability to intersect the Solar System \cite{Kistler:2012ag}.

Concerning the low energy part of the CRE spectra, where heliospheric propagation effects are relevant, we use the recently developed {\sc HelioProp} code \cite{Maccione:2012cu} which solves the CR propagation equation in the Solar System accounting for charge-dependent drifts (see also \cite{Bobik:2011ig}).

In this way, we describe consistently the CR fluxes at Earth exploiting 3D propagation models on both galactic and solar system scales.

{\em Propagation setup.} 
The CR transport in the Galaxy is described by the well-known transport equation \cite{1964ocr..book.....G,1965P&SS...13....9P}.
For each CR particle, we solve the set of coupled transport equations with the numerical code {\sc DRAGON} \cite{Dragonweb,Evoli:2008dv,DiBernardo:2010is}, suitably extended to describe generic 3D spatial geometries. The code can deal with arbitrary CR source and gas distributions, generic magnetic field models, and fully anisotropic, position-dependent diffusion. 
For our purposes, we can take the diffusion coefficient as a scalar with the following dependences on the rigidity $\rho$: 
$D(\rho) = \beta^{-0.4}~ D_0 \left( \rho/\rho_0 \right)^\delta$, with $\rho_{0}=3~\GV$ \cite{DiBernardo:2010is}. 
\begin{figure}[tbp]
\centering
\includegraphics[width=0.4\textwidth]{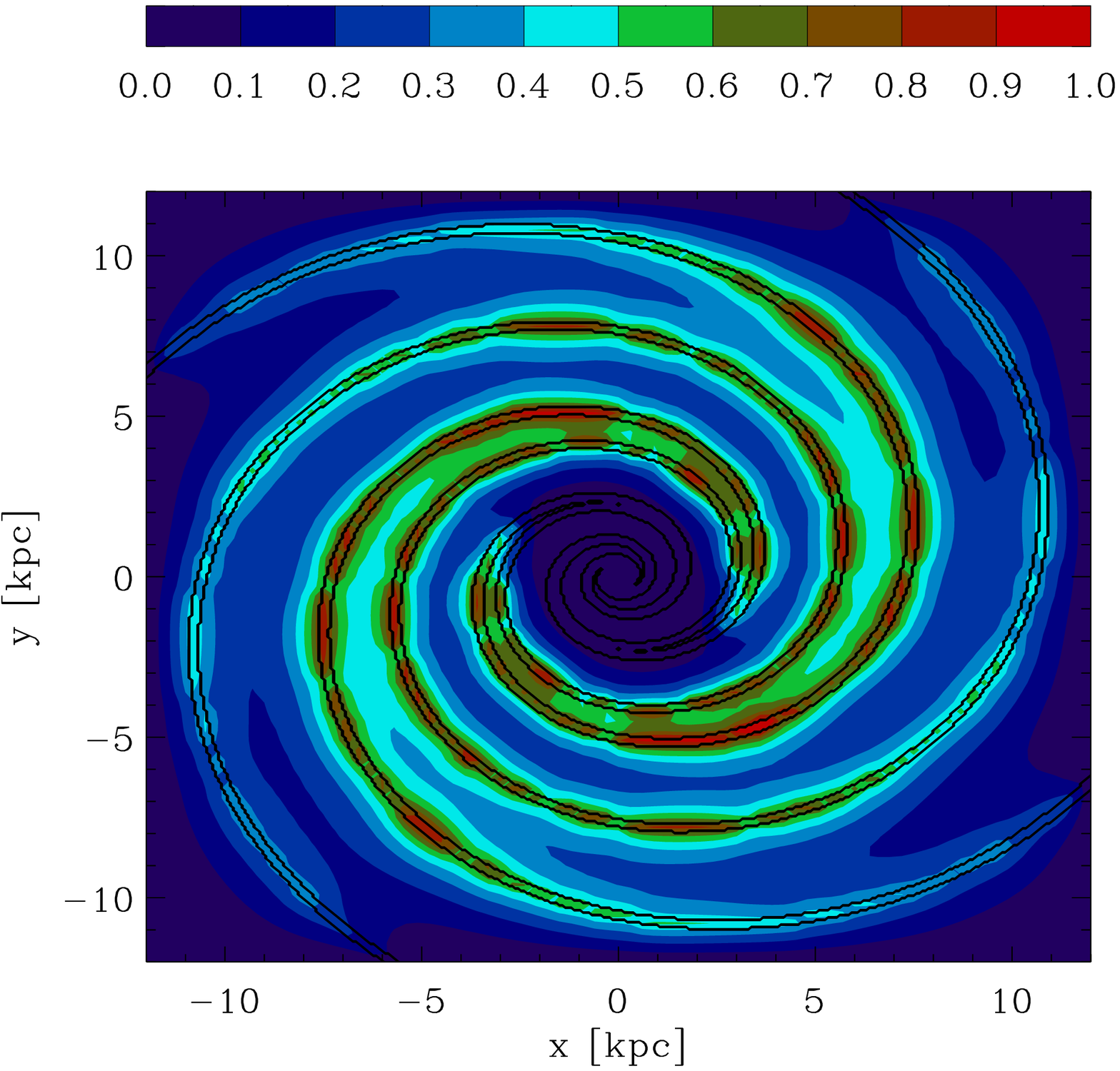}
\includegraphics[width=0.4\textwidth]{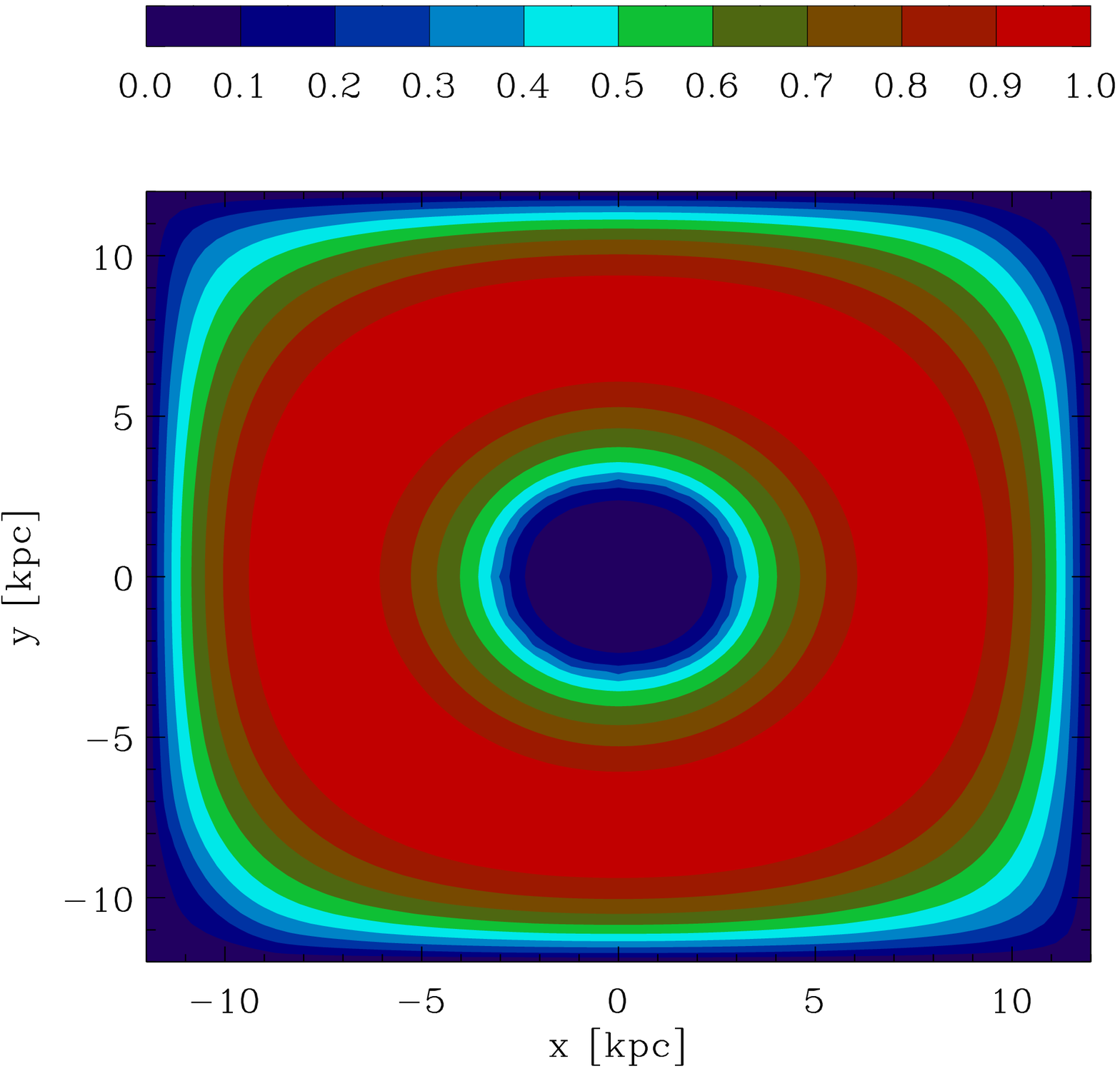}
\caption{Top view of the propagated distribution, normalized to its maximum, on the Galactic plane of $e^{-}$ at 100~GeV for sources distributed in the spiral arms (top panel) or smoothly (bottom panel). In the first case the contour of the assumed source distribution is superimposed (black lines).}
\label{fig:spirals}
\end{figure}

We consider a single benchmark propagation setup characterized by the following parameters: $D_{0}=3\times10^{28}~\cm^{2}/\s$, $\delta=0.6$, half-halo height of $4~\kpc$, and no reacceleration. 
For all nuclear species we use an unbroken momentum power-law source spectrum with the same spectral index $\gamma_{0, p} = 2.28$ (we checked that the propagated proton spectrum is in agreement with the one derived in \cite{Dermer:2013iwa,Loparco:2013pea}).   
For the $e^-$ {\em background} source spectrum we assume a broken power law \cite{DiBernardo:2012zu,Strong:2011wd}: below $4~\GeV$ we adopt a spectral index $1.2$ (slightly different from the 1.55 found in \cite{Potgieter:2013xhj}, which, however, is hardly consistent with the galactic diffuse synchrotron emission spectrum \cite{DiBernardo:2012zu}) while above that energy we tune it against PAMELA data (see below). For the interstellar radiation field we use the model \cite{Porter:2005qx}.

This model provides a very good combined fit of B/C and proton spectra, respectively, measured by HEAO-3 \cite{1989ApJ...346..997B} and PAMELA \cite{Adriani:2011cu} and also matches $^4$He and other nuclear species' absolute and relative spectra provided the solar modulation treatment described in \cite{Maccione:2012cu} is consistently used. We remark that different choices of the relevant parameters would not significantly affect our results, provided that all the above data sets are consistently reproduced. 
We verified that introducing a spectral break at $E \sim 200~\GeV$ in order to reproduce the spectral hardening of the proton and helium spectra observed in PAMELA and CREAM \cite{Ahn:2010gv} measurements does not have a significant effect on the positron and electron spectra in the energy range we consider in this work. 

With the new 3D code, we can release the azimuthal symmetry hypothesis and introduce a realistic spiral arm pattern for the source distribution. Here we adopt the model used in \cite{Blasi:2011fi} (see also \cite{Pohl:1998ug,Mertsch:2009ph}), which reproduces the observed spiral structure of the Milky Way. We show in Fig.~\ref{fig:spirals} a face-on view of the density of propagated CR primary $e^{-}$ on the Galactic plane at 100 GeV, together with the distribution of their sources, for the two cases with and without the spiral arm distribution. The effect of the source distribution is striking: In the spiral arm case the electrons are more closely attached to their parent arm, while in the other case they are more uniformly spread. We verified explicitly that the spiral arm structure does not affect significantly the spectrum of protons and nuclei, nor that of low energy ($\lesssim20~\GeV$) secondary and primary electrons and positrons, as expected because of the longer mean-free path of these particles with respect to high energy leptons. 

We treat solar modulation with the recently developed {\sc HelioProp} \cite{Maccione:2012cu}. Before they are detected at Earth, CRE lose energy due to the solar wind while diffusing in the Solar System \cite{Gleeson_1968ApJ}.  Because of drifts in the large scale gradients of the solar magnetic field (SMF), the modulation effect depends on the particle charge including its sign \cite{1996ApJ...464..507C}. Therefore, it depends on the polarity of the SMF, which changes periodically every $\sim$11 years \cite{wilcox}. 

The SMF also has opposite polarities in the northern and southern hemispheres: At the interface between opposite polarity regions, a heliospheric current sheet (HCS) is formed (see e.g.~\cite{1981JGR....86.8893B}). The HCS swings then in a region whose angular extension is described phenomenologically by the tilt angle $\alpha$. Its magnitude depends on solar activity. Since particles crossing the HCS suffer from additional drifts because of the different orientation of the magnetic field lines, the intensity of the modulation depends on the extension of the HCS \cite{2012Ap&SS.339..223S}. Besides $\alpha$, another important parameter is related to diffusion. 
As in \cite{Maccione:2012cu}, we assume that diffusion occurs in the Bohm regime, and that its intensity is described as $D(\rho)=\lambda(\rho) v/3$, with $\lambda$  the momentum-dependent mean-free path. 

{\em Results.}  We first assume the extra-component sources to be located in the spiral arms and have the same distribution as standard SNRs and disregard, for the moment, the possible role of local sources.  
\begin{figure}[tbp]
\centering
\includegraphics[width=0.45\textwidth]{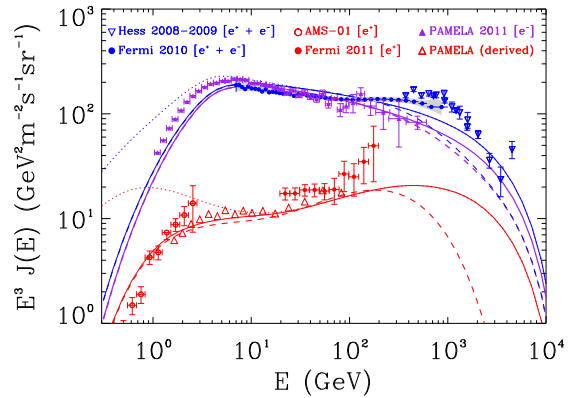}
\caption{ The $e^- + e^+$ (blue), $e^-$ (purple) and $e^+$ (red) propagated spectra computed in our model.  Solid (dashed) lines are for the case of a SNR- (pulsar-) like contribution.  Dotted lines are for the interstellar spectra. PAMELA $e^+$ data have been derived (without error propagation) starting from the PF and $e^-$ spectrum released by the same collaboration. We warn the reader that this derivation might be subject to large systematics, especially below $\sim20~\GeV$, because the $e^{-}$ and the PF datasets were taken in different periods.}
\label{fig:spirals_onlyelpos}
\end{figure}
\begin{figure}[tbp]
\centering
\includegraphics[width=0.45\textwidth]{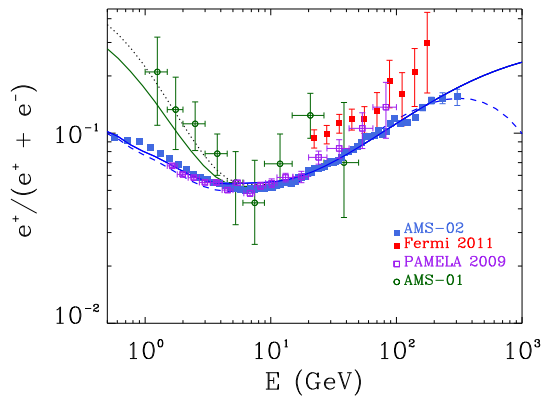}
\caption{PF computed in our model. The blue (green) curves correspond to the AMS-02 (AMS-01) data taking periods. The solid (dashed) curves are for a SNR- (pulsar-) like contribution at high energy. The dotted line is the interstellar PF.}
\label{fig:spirals_onlyPF}
\end{figure}
For the extra-component source spectrum we assume the form $J_{EC}(e^\pm) \propto E^{-\gamma_{0, {\rm EC}}}  \exp({-E/E_{\rm cut}})$ and tune the involved parameters against the data. We normalize the primary electron component to the PAMELA $e^{-}$ flux at 33~GeV. In these conditions the AMS-02 PF and the PAMELA and Fermi LAT $e^-$ spectra can consistently be reproduced if $\gamma_{0, {\rm EC}} \simeq 1.75$ and $E_{\rm cut} \simeq 10~\TeV$
(see Figs.~\ref{fig:spirals_onlyelpos} and \ref{fig:spirals_onlyPF}). Remarkably, passing from a smooth source distribution to a more realistic spiral arm pattern, a {\em harder} $e^-$ source spectral index is required: $\gamma_{0, {\rm bkg}}\simeq2.38$, to be contrasted with $\gamma_{0, {\rm bkg}}\simeq 2.65$ used, e.g., in \cite{DiBernardo:2010is}. As we already pointed out, this is a consequence of the Solar System being placed between two main arms (Perseus and Sagittarius-Carina), hence in a source underdense region. This turns into a larger average distance, hence stronger losses, between the bulk of sources in the arms and the observer. 
The $e^+$ spectrum measured by AMS-01 and that computed on the basis of PAMELA PF and $e^-$ spectrum (preliminary  PAMELA $e^+$ results agree with this estimate) are nicely matched by our model.  
The spectral steepening found by H.E.S.S. \cite{Aharonian:2009ah} is also naturally reproduced with a very high energy cutoff as that expected in the scenario envisaged in \cite{Blasi:2009hv,Mertsch:2009ph,Ahlers:2009ae} where $e^{\pm}$ are produced as secondaries of CR nuclei and reaccelerated in SNRs.
Pulsars would hardly provide those high energies. The pulsar scenario, however, cannot be excluded on the basis of this data. Indeed, an alternative model also compatible with the pulsar scenario, with a lower cutoff energy $E_{\rm cut} \simeq 1~\TeV$ and a slightly harder spectral index $\gamma_{0, {\rm EC}} \simeq 1.5$, can reproduce the data comparably well, as also shown in Figs.~\ref{fig:spirals_onlyelpos} and \ref{fig:spirals_onlyPF}. In this latter case, the PF is expected to flatten out and decrease above $\sim 300~\GeV$. Remarkably, observations of the PF at these high energies could distinguish the two models.

\begin{figure}[tbp]
\centering
\includegraphics[width=0.45\textwidth]{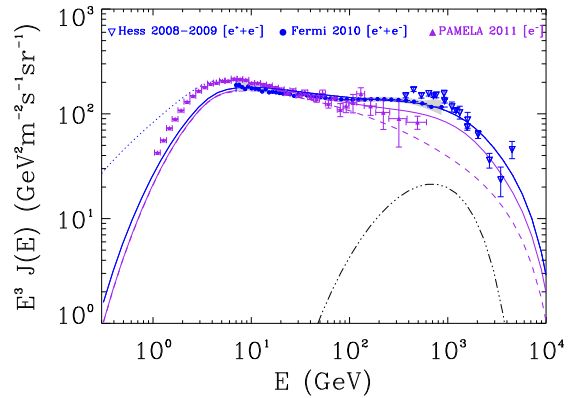}
\caption{The $e^- + e^+$ (solid blue), and $e^-$ (solid purple) propagated spectra computed assuming that one nearby electron accelerator is also present (triple-dot-dashed curve). The dashed curve represents the background $e^{-}$ component, while the dotted curve is for the interstellar spectrum.}
\label{fig:Fermielpos}
\end{figure}

\begin{figure}[tbp]
\centering
\includegraphics[width=0.45\textwidth]{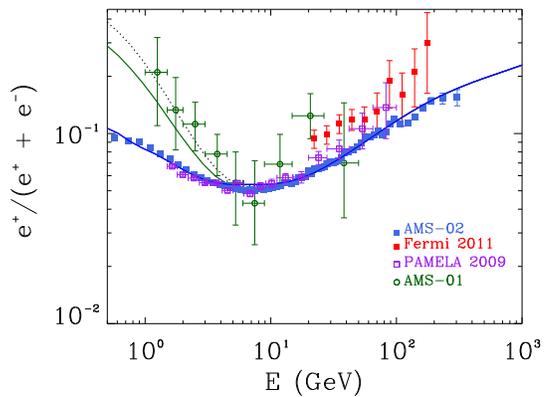}
\caption{The PF under the same hypothesis as Fig.~\ref{fig:Fermielpos}. The line notation is the same as Fig.~\ref{fig:spirals_onlyPF}.}
\label{fig:FermiPF}
\end{figure}

The CRE spectrum measured by Fermi LAT is not satisfactorily reproduced. This discrepancy can point to the presence of unknown systematics in Fermi LAT data. However, it could also be a signal of the emergence of a nearby $e^{-}$ source at high energy. We show in Figs.~\ref{fig:Fermielpos} and \ref{fig:FermiPF} the CRE and $e^{-}$ spectra and the PF computed for the same model used above, but normalized this time to the CRE flux of Fermi LAT at 33~GeV. 

While the PF is correctly reproduced, the CRE spectrum above $\sim$200~GeV is underproduced. This leaves room for a possible contribution of $e^{-}$ by a local source at high energies. This is shown as the triple-dot-dashed curve in Fig.~\ref{fig:Fermielpos}. The injection spectrum of this component is a power law with exponential cutoff, with $\gamma_{0{\rm , L-EC}} = 2.1$ and $E_{\rm cut, L-EC} = 1~\TeV$. The required energy output is $3.6\times10^{47}~\erg$ if the source is located at a distance $d\simeq290~\pc$ from the Solar System, compatible with the position of the Vela SNR. The dipole anisotropy due to this single source is diluted in the sea of the Galactic extra component and is $\sim0.8\%$ at 100~GeV, reaching a maximum of $\sim 2~\%$ at 500~GeV, which is significantly smaller than present experimental upper limits (see also \cite{Linden:2013mqa} for a discussion of the role of anisotropy in view of the CTA \cite{Vandenbroucke:2011sh} in the local source scenario).  

At low energy, we compute our spectra using the solar propagation parameters appropriate for the data-taking period of each dataset. In particular, we use for AMS-01 $\alpha\simeq10^{\circ}$ and positive polarity, while for PAMELA and AMS-02 we use $\alpha = 10^{\circ}$ and $\alpha=60^{\circ}$, respectively, and negative polarity. We tune on the data the  remaining free parameter $\lambda$. Remarkably, we also achieve a very good fit of all PF data sets in both our models (see Figs.~\ref{fig:spirals_onlyPF} and \ref{fig:FermiPF}) by tuning $\lambda=0.2$ and $\lambda=0.4$ at $1~\GeV$ for AMS-02 and both PAMELA and AMS-01, respectively. These values are in rough agreement with findings from the analysis of the time dependence of proton spectra \cite{Potgieter:2013cwj}. We checked that this model reproduces the spectrum of the Galactic diffuse radio emission between 10~MHz and 3~GHz (see \cite{DiBernardo:2012zu} for more details).

{\em Conclusions.} We have computed the CR electron and positron spectra at Earth within a 3D numerical propagation model. For the first time, we have computed the consequences of the CR sources being mainly distributed in spiral patterns in the Galaxy, with the Solar System lying in an interarm region. 
The resulting picture is, for the first time, compatible with expectations from shock acceleration scenarios. We will consider in a forthcoming work even more realistic models for the Galaxy. Many features may contribute to slightly enhance or reduce the effects we studied, e.g.~the halo height, the distribution of interstellar radiation and magnetic fields, and the arm-interarm contrast for the source distribution (our model is equivalent to a Gaussian profile for the spiral with a full width at half maximum of $\sim$900~pc).

At high energies, PAMELA and AMS-02 observations are compatible if the extra component is charge symmetric and sources are distributed in the spiral arms. SNRs as envisaged in \cite{Blasi:2009hv,Mertsch:2009ph,Ahlers:2009ae} or pulsars \cite{Hooper:2008kg}, may be viable source candidates.  Measurements of the PF at $\sim500~\GeV$ could distinguish the two possibilities. This scenario naturally avoids tension with the smoothness of the observed spectra and with constraints on dipole anisotropies. However, Fermi LAT data are not well reproduced in this model. Barring unaccounted systematics, compatibility with Fermi LAT data can be obtained by adding a contribution of only electrons as it may be expected from a local SNR (e.g.,~Vela). 

At low energies, we used charge-dependent solar modulation to successfully reproduce data taken in different periods of solar activity. This further strengthens the need for an accurate description of solar propagation to interpret data below 10 GeV.

{\em Acknowledgments.} D.~Gaggero warmly thanks the Max-Planck-Institute for Physics for hospitality during the preparation of this work and P.~Ullio for useful discussions. L.M. acknowledges support from the Alexander von Humboldt foundation and partial support from the European Union FP7 ITN INVISIBLES (Marie Curie Actions, PITN-GA-2011-289442). C.E. acknowledges support from the "Helmholtz Alliance for Astroparticle Physics HAP" funded by the Initiative and Networking Fund of the Helmholtz Association. We thank I.~Gebauer and the Karlsruher Institut f\"ur Technologie (KIT) for providing computing resources and precious help in code debugging. We also thank P.~D.~Serpico for useful comments on a preliminary version of this Letter.

\bibliographystyle{apsrev4-1}
\bibliography{AMS_letter}
    
\end{document}